# Wikipedia in academia as a teaching tool: from averse to proactive faculty profiles

J. Minguillón, E. Aibar, M. Lerga, J. Lladós, A. Meseguer-Artola

Universitat Oberta de Catalunya, Barcelona, Spain
Rambla Poblenou 156, Barcelona 08018, Spain
jminguillona@uoc.edu

**Abstract**

***Introduction***. *This study concerned the active use of Wikipedia as a teaching tool in the classroom in higher education, trying to identify different usage profiles and their characterization.*
***Method***. *A questionnaire survey was administrated to all full-time and part-time teachers at the Universitat Oberta de Catalunya and the Universitat Pompeu Fabra, both in Barcelona, Spain. The questionnaire was designed using the Technology Acceptance Model as a reference, including items about teacher's web 2.0 profile, Wikipedia usage, expertise, perceived usefulness, easiness of use, visibility and quality, as well as Wikipedia's status among colleagues and incentives to use it more actively.*
***Analysis***. *Clustering and statistical analysis were carried out using the k-medoids algorithm and differences between clusters were assessed by means of contingency tables and generalized linear models (logit).*
***Results***. *The respondents were classified in four clusters, from less to more likely to adopt and use Wikipedia in the classroom, namely averse (25.4%), reluctant (17.9%), open (29.5%) and proactive (27.2%). Proactive faculty are mostly men teaching part-time in STEM fields, mainly engineering, while averse faculty are mostly women teaching full-time in non-STEM fields. Nevertheless, questionnaire items related to visibility, quality, image, usefulness and expertise determine the main differences between clusters, rather than age, gender or domain.*
***Conclusions***. *Clusters involving a positive view of Wikipedia and at least some frequency of use clearly outnumber those with a strictly negative stance. This goes against the common view that faculty members are mostly sceptical about Wikipedia. Environmental factors such as academic culture and colleagues' opinion are more important than faculty's personal characteristics, especially with respect to what they think about Wikipedia's quality.*

**Introduction**

Wikipedia is one of the most successful examples of a massive collaborative effort made possible thanks to the birth of the Web, receiving a lot of attention from the scientific community (Okoli et al., 2014). With millions of entries in hundreds of different languages, it has become an immediate source of information for almost any knowledge topic. Anyone can contribute by adding, editing and improving articles according to their preferences and expertise. As described by Samoilenko and Yasseri (2014), Wikipedia is 'a web-based encyclopaedia which allows any user to freely edit its content, create and discuss the articles - all in the absence of central authority or stable membership. This model of a decentralized bottom-up knowledge construction draws on the wisdom of the crowds rather than on professional writers and peer-reviewed material, which makes Wikipedia similar to a social media platform. Unlike other encyclopaedias, Wikipedia is unrestricted in size and range of topics covered, and thereby holds the potential to become the most comprehensive repository of human knowledge.'

Wikipedia represents the junction of open educational resources and Web 2.0 tool initiatives. On the one hand, it has become a gigantic open repository of knowledge and information with great potential for use in learning processes at different levels of education (Konieczny, 2014; Nix 2010). On the other hand, it can be considered a primary example of so-called mass-online commons-based peer production (Benkler, 2006; Weber, 2004), where the collective construction of knowledge takes place through a virtual platform that facilitates collaboration on an unprecedented scale.

One of the biggest paradoxes in Wikipedia is that higher education faculty, who are supposed to be experts in their domain of knowledge, are thought to be reluctant to use and recognise Wikipedia as a valuable teaching tool (Jashchick, 2007; Brox, 2012; Bayliss, 2013). In the university context, this open peer production (Ricaurte-Quijano and Alvarez, 2016) is at possible odds with science due to authorship dilution and a replacement of the typical expert-based peer-reviewed process as a mechanism for quality control. However, Wikipedia is one of the resources most employed by students, who use it regularly as a reference tool and to carry out different assignments and tasks (Brox, 2012; Knight and Pryke, 2012). According to Lim (2009), students value not just the quality of Wikipedia's articles but also the easy access to its content, the hypertext structure that facilitates navigation, and the abundance of references and sources.

Faculty attitudes seem less positive, and more sceptical. Despite a positive overall assessment of the information's quality, there are concerns about how students are using Wikipedia as a source of information (Chen S., 2010; Bayliss, 2013; Chandler and Gregory, 2010; Traphagan et al. 2014). In fact, many academics discourage citing Wikipedia as a source, because their articles do not have a clear and identifiable authorship, which therefore makes it difficult to verify their content (Jaschick, 2007; Santana and Wood, 2009). Moreover, some faculty consider that Wikipedia demotivates students from using other, more reliable, sources of information (Dooley, 2010). Credibility in Wikipedia with respect to its actual use has been reviewed and analyzed by Francke and Sundin (2010). Wikipedia seems to be a good starting point for any search for information about any topic, but users are hesitant to use or refer to it in situations where they need to be certain of something. In a similar vein, H. Chen (2010) identified (lack of) credibility as university faculty's main concern about Wikipedia. These negative views are generally embedded in academia, but their strength depends on scientific discipline, being stronger in soft disciplines, including social sciences, arts and humanities (Kemp and Jones, 2007; Ehmann et al, 2008; Smith, 2012), rather than in hard ones, or STEM (Science, Technology, Engineering and Math degrees).

Other possible explanations have to do with Wikipedia's particular way of producing and assessing knowledge contents, something closely linked to the peer production model it represents (Benkler, 2006). In general, Wikipedia has been the subject of many academic studies (Okoli et al., 2014). Some authors relate accuracy and credibility concerns to a more fundamental conflict on epistemological and power-struggle grounds (Black, 2008; S. Chen, 2010; Eijkman, 2010). As stated by Gray et al. (2008), the use of Wikipedia redefines the boundaries of scholarly and scientific communication in several aspects, including formal (authorised) and informal (unofficial) publications. There is also a growing interest in the role of Wikipedia as a channel for the public communication of scientific outputs (Brossard and Scheufele, 2013; Halfaker and Taraborelli, 2015; Nielsen, 2007) as well as a teaching tool (Christensen, 2015; Konieczny, 2016).

Unfortunately, empirical studies on faculty perceptions and uses of Wikipedia in learning environments are few and quite limited in scope. In relation to consultation, Snyder points out that faculty mainly use Wikipedia for independent learning or leisure (almost 80% of respondents) while less than 10% uses it as a source of information when writing an academic paper (Snyder, 2010). In a similar vein, Xiao and Askin (2012) focus on using Wikipedia for academic publishing and state that almost 60% of respondents had no experience with

Wikipedia besides that of a reader. Only (Aibar et al., 2015) asked about proactive teaching use of Wikipedia among faculty members, with a positive result of less than 10%. The authors explore relationships between faculty perceptions and several characteristics. Not only do they establish the extent to which sceptical attitudes are related to disciplinary or generational factors, they also unearth an implicit conflict between the standard scientific or academic epistemological stances and the specific peer-to-peer culture of Wikipedia (as a paradigmatic example of content production in a collaborative open network) (Aibar et al., 2015).

The educational benefits of using Wikipedia in higher education as a teaching tool have been described from a qualitative perspective. An and Williams (2010) mention some of them: fostering of interaction; communication and collaboration among students; improvement in writing and technological skills; the ease of use and flexibility; and a new role of faculty as facilitators of learning rather than distributors of knowledge. Maehre (2009) remarks on the benefits of using Wikipedia for the students' information literacy and their critical thinking, in addition to students learning by engaging in a process instead of creating a product. Finally, Kennedy et al. (2015) propose using Wikipedia for both improving student learning outcomes and benefit a much broader community. From a practical point of view, several authors have also explored the use of Wikipedia as a teaching tool and have identified several educational benefits (Forte and Bruckman, 2006; Konieczny, 2007; Wannemacher, 2009; Konieczny, 2012; Meseguer-Artola, 2014). For instance, Wikipedia can be used to learn a foreign language, to write an essay about any topic and even as a huge source of data for social network analysis. On the other hand, Wikipedia promotes collaborative work as well as the acquisition and development of informational competences, which are known to be 21st century skills (Christensen, 2015; Konieczny, 2016).

Nevertheless, there is a lack of studies about the underlying reasons that make faculty to use Wikipedia as a teaching tool, or promote it among their students and colleagues. Although it is an individual decision, it is probably shaped by academic environment (Xiao and Askin, 2014) and the way innovative teaching practices are adopted by faculty, according to their scientific discipline (Smith, 2012).

Our main research question is to check, based on a large empirical study, whether most faculty members actually share a clear negative opinion on Wikipedia or not, as the usual view has it, particularly in the context of higher education. We undertake this task by means of a cluster analysis that identifies intermediate stands and not only extreme ones. Then, as a secondary research question, we explore what elements are relevant in characterising the clusters found and thus can be taken as driving factors in shaping attitudes towards Wikipedia.

**Survey**

The main tool used for data gathering was an online survey sent to all faculty members of the Universitat Oberta de Catalunya (UOC), in the first wave, and to Universitat Pompeu Fabra (UPF) faculty members later on, in a second wave. Both universities are located in Spain. The UOC is a completely online university, launched in 1994, that uses ICTs extensively to provide its community with a comprehensive virtual campus where all the teaching and learning activities are carried out. With more than 60,000 students, the faculty comprises (approximately) 250 full-time teachers and almost 2,000 part-time associate teachers, all of them considered faculty members. The UPF was launched in 1990 and it is Catalonia's youngest public university. It has around 12,000 students and 1,511 faculty members. Previously to the survey, an exploratory qualitative study was carried out to collect data in order to improve its design. Twelve interviews with faculty members were conducted, selecting two from each of the six main schools at UOC. These interviews were conducted between 12 and 16 October 2012.

In the first wave, from a universe of 2,128 individuals, 800 valid responses were received. For a confidence level of 95%, and the assumption of maximum uncertainty ($p = q = 0.5$), the margin of error was 2.74%. The second wave was addressed to a universe of 1,511 individuals and 113 valid responses were received (confidence level of 95%, $p = q = 0.5$ and margin of error = 8.87%). As shown in Table 1, 913 valid responses provided the final sample size, from a universe of 3,639 individuals. A complete description of the survey can be found in (Aibar et al., 2015).

| | |
|---|---|
| Study universe | Faculty members of the Universitat Oberta de Catalunya and the Universitat Pompeu Fabra |
| Study universe size | 3639 |
| Method | Online survey sent to the universe, with no quota groups |
| Sample size | 913 |
| Sampling error | ±2.81% for overall data in the case of maximum uncertainty (p=q=0.5). Confidence level 95%. |
| Resulting sample | Not weighted |
| Date of launching | November 19$^{th}$, 2012 (UOC) and April 29$^{th}$, 2013 (UPF) |
| Data collection | From November 19$^{th}$ to December 3$^{rd}$, 2012 (UOC) and April 29$^{th}$ to May 16$^{th}$, 2013 (UPF) |

**Table 1: Technical information on the questionnaire.**

Since we consider Wikipedia to be a technological platform for knowledge creation and sharing, the technology acceptance model (TAM) was used to study how faculty members adopt and use Wikipedia, according to the definition of Bagozzi, Davis and Warshaw (1992). In (Meseguer-Artola et al., 2016), the authors describe a conceptual model as an adaption of the so-called TAM 3 model, where special attention is paid to the external factors that influence the acceptance of a particular technology (Venkatesh and Bala, 2008). This is a more comprehensive vision of the traditional model, which considers that experience might affect behavioural intention and perceived ease of use. In this model, the following 6 factors are considered, as defined in (Meseguer-Artola et al., 2016): job relevance, sharing attitude, social image, profile 2.0, quality of Wikipedia, and perceived enjoyment. With this approach the authors were able to study (a) how faculty make a decision about the adoption and use of Wikipedia, and (b) what external variables and prior factors can lead to greater acceptance and effective utilization of Wikipedia.

The questionnaire was organized into two parts. The first part was aimed at collecting data on gender, age group, area of expertise (STEM or non-STEM), PhD degree, years of experience in university teaching according to Huberman's scale (1993), academic position (adjunct faculty working part-time or part-time or full-time academic faculty) and registered membership in Wikipedia. The second part, with 41 questions, was designed to measure different thoughts and attitudes towards Wikipedia (Aibar et al., 2015), mainly its perceived quality, teaching practices involving Wikipedia, usage experience, perceived usefulness and use of 2.0 tools, using a 5-point Likert scale, measuring the level of agreement or disagreement with a statement (1="Strongly disagree" and 5="Strongly agree") or to the frequency of certain actions (1="Never" and 5="Very often"). The questions that were related to the active uses of Wikipedia as a teaching tool in the classroom are the following:

- USE1: I use Wikipedia to prepare my teaching materials.
- USE2: I use Wikipedia as a teaching platform for developing activities with my students.
- USE3: I recommend that my students use Wikipedia.
- USE4: I recommend that my colleagues use Wikipedia.
- USE5: I let my students use Wikipedia.

Notice that the first two questions (USE1 and USE2) refer to proactive uses of Wikipedia as a teaching tool. The other three questions (USE3, USE4 and USE5) refer to whether faculty promote or discourage the use of Wikipedia among their students and other colleagues. We were interested in establishing a relationship between proactively using Wikipedia and spreading the fact of doing so, according to a TAM model (Meseguer-Artola et al., 2016).

**Method**

Clustering techniques try to group similar elements in homogeneous clusters, while separating different elements and identifying distinct characteristics among them. In our case, each faculty member was located in a cluster according to her answers to the given construct (i.e. the selected subset of questions from the survey). With this process, we expected to uncover the array of existing faculty behaviours with respect to Wikipedia use as well as discover the different profiles populating each cluster.

Before proceeding with the clustering algorithm, we needed to check the reliability of the proposed construct. For USE1 to USE5, Cronbach's α was 0.87, which is considered to be good (Hair et al., 2010). An unrotated principal component analysis (PCA) revealed a unique component with eigenvalue > 1 that explained 67.5% of total variance, with item loadings ranging from 0.39 to 0.50. Hence, we concluded that these items sufficiently captured the use and acceptance of Wikipedia as a teaching tool.

We only used USE1 to USE5 as the inputs for the clustering algorithm. As these variables are ordinal, we chose to use the medoids version of the k-means algorithm (k-medoids), that is, the centroid for each one of the computed clusters will always be an element from the original data set (Kaufman and Rousseeuw, 1987). This allowed us to better understand the nature of each cluster found, and its members. We used the PAM algorithm implemented in R and the Gower (1971) dissimilarity index as the distance measure. The fact that the five variables were ordinal (i.e. the distance from 1 to 2 does not necessarily mean the same as the distance from 2 to 3) justified the use of a non-Euclidean distance criterion.

One of the problems with the k-medoids algorithm was the need to specify the number of clusters (K) a priori, that is, the algorithm provided no clue to discovering the optimal number of clusters, which it needed as an input parameter. Nevertheless, as the number of observations and variables was small, it was perfectly feasible to use brute force to compute different clusterings using K=2, K=3, and so on. The optimal number of clusters was then determined using the Calinski-Harabasz (1974) criterion. As shown in Table 2, a maximum could be found at K=4.

| K | 2 | 3 | 4 | 5 | 6 | 7 | 8 | 9 | 10 |
|---|---|---|---|---|---|---|---|---|---|
| CH | 995.7 | 857.6 | 1008.2 | 990.8 | 878.8 | 814.9 | 749.9 | 709.4 | 720.5 |

Table 2: Computed Calinski-Harabasz criterion (CH) according to the desired number of clusters.

**Results**

The four clusters obtained by the k-medoids algorithm are shown in Table 3. For each cluster, we show the number of elements in the cluster (N), the values of the five variables USE1 to USE5 that determine the medoid, namely the median, mean and standard deviation and the name used to identify the cluster.

| Cluster | N | USE1 | USE2 | USE3 | USE4 | USE5 | Name |
|---|---|---|---|---|---|---|---|
| 1 | 253 | 2<br>1.92<br>0.83 | 1<br>1.44<br>0.71 | 3<br>3.04<br>0.67 | 3<br>2.88<br>0.74 | 4<br>3.73<br>0.80 | open |
| 2 | 218 | 1<br>1.23<br>0.56 | 1<br>1.09<br>0.33 | 1<br>1.21<br>0.44 | 1<br>1.21<br>0.45 | 2<br>2.24<br>0.95 | averse |
| 3 | 233 | 3<br>3.32<br>0.87 | 3<br>2.95<br>1.04 | 4<br>4.00<br>0.67 | 4<br>3.83<br>0.76 | 4<br>4.07<br>0.75 | proactive |
| 4 | 153 | 2<br>1.84<br>0.70 | 2<br>1.82<br>0.80 | 2<br>2.01<br>0.38 | 2<br>1.94<br>0.43 | 3<br>2.95<br>0.71 | reluctant |

**Table 3: Computed clusters obtained with the k-medoids algorithm (K=4). Each cell shows median, mean and standard deviation for each variable.**

Cluster 1, the most populated, reproduces (not surprisingly) typical behaviour with respect to Wikipedia uses: faculty use Wikipedia as a source for preparing their classrooms, but they do not use it as a learning platform with their students, although they think it is fine to use it. We have named this cluster open. Cluster 2, named averse, includes those faculty members that do not use Wikipedia and do not promote its usage among students and colleagues. Cluster 3 (named proactive) includes those faculty members that proactively use and promote Wikipedia. Finally, Cluster 4 (named reluctant) includes those faculty members falling between the averse and proactive clusters, with a relatively low use of Wikipedia.

In order to improve the visualisation of the differences among clusters, we have sorted the clusters according to their meaning, rather than their index. Therefore, from less proactive to more proactive, we show clusters 2, 4, 1 and 3, respectively. Figure 1 shows a multidimensional scaling plot (Borg and Groenen, 2005) using the same distance matrix computed for clustering purposes. V1 and V2 have no special meaning; they are just computed for visualisation purposes. The medoid for each cluster is represented using a larger graphical element. Note that clusters 2 and 3 are perfectly separable (i.e. they do not overlap); clusters 4 and 1 are in between; cluster 4 is slightly closer to cluster 2; and cluster 1 is closer to cluster 3.

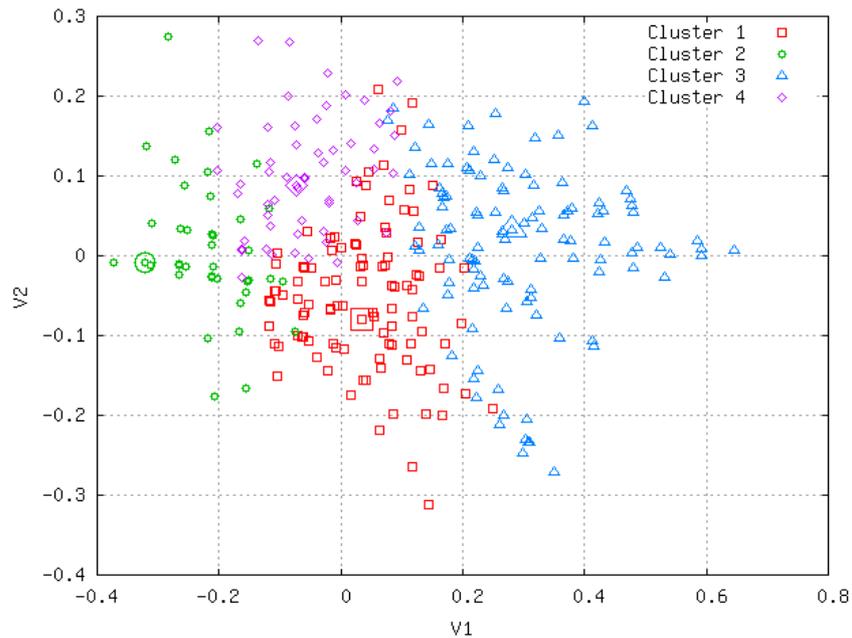

Figure 1: Multidimensional scaling of USE1 to USE5 variables using Gower's dissimilarity index.

As mentioned, the four clusters seem to reflect the variety of users in terms of their usage of Wikipedia for teaching purposes. Preliminary experiments (using only data from UOC faculty) showed that control variables such as gender and domain are important, but their effects vanish when other items related to user attitudes and perceptions are included (Meseguer-Artola et al., 2016). We characterised the users in each one of the clusters and compared whether there were significant statistical differences among them or not. Firstly, we used contingency tables to test the differences between clusters for each variable. Then we built generalised linear models to explore interactions between variables.

Table 4 shows the basic profiling variables used to characterise the users in each cluster. For binary variables, mean is the probability of satisfying the condition described by the name of the variable (i.e. being female), that is, a proportion. The mean for each cluster is also shown; an asterisk ("*") indicates that there are standardised residuals with an absolute value larger than 3 for said cluster, which indicates that there is a strong relationship between the cluster and at least one of the possible values taken by the variable. For instance, the asterisk for the combination Female/Cluster 3 indicates that the proportion of females in this cluster is significantly different (> 3 standardised sigma) from the expected proportion of females.

| Variable | N | | Mean | Cluster 2 | Cluster 4 | Cluster 1 | Cluster 3 | p value |
|---|---|---|---|---|---|---|---|---|
| Gender MALE=0 FEMALE=1 | 857 | 499 | 0.5823 | 0.4633 | 0.5229 | 0.5968 | 0.7167 | < 0.001 |
| | | 358 | 0.4177 | 0.5367 | 0.4771 | 0.4032 | 0.2833* | |
| Age | 857 | | 42.06 | 41.98 | 42.22 | 41.88 | 42.21 | 0.9054 |
| Expertise <= 6 years=0 > 6 years=1 | 836 | 296 | 0.3541 | 0.3160 | 0.4094 | 0.3618 | 0.3450 | 0.3234 |
| | | 540 | 0.6459 | 0.6840 | 0.5906 | 0.6382 | 0.6550 | |
| Faculty ADJUNT=0 ACADEMIC=1 | 850 | 466 | 0.5482 | 0.4700 | 0.5490 | 0.5378 | 0.6332 | 0.0069 |
| | | 384 | 0.4518 | 0.5300 | 0.4510 | 0.4622 | 0.3668 | |
| PhD NO=0 YES=1 | 857 | 461 | 0.4621 | 0.5000 | 0.4706 | 0.4901 | 0.3906 | 0.0752 |
| | | 396 | 0.5379 | 0.5000 | 0.5294 | 0.5099 | 0.6093 | |
| Registered NO=0 YES=1 | 855 | 739 | 0.8643 | 0.9725 | 0.8889 | 0.8814 | 0.7273 | < 0.001 |
| | | 116 | 0.1357 | 0.0275* | 0.1111 | 0.1186 | 0.2727* | |
| STEM NO=0 YES=1 | 856 | 655 | 0.7652 | 0.8670 | 0.8105 | 0.7589 | 0.6466 | < 0.001 |
| | | 201 | 0.2348 | 0.1330* | 0.1895 | 0.2411 | 0.3534* | |

Table 4. Differences of means among profiling variables used to characterise users.

Notice that, in most cases, clusters 2 and 3 display very opposite behaviours, while clusters 4 and 1 have no particular characteristic. In fact, according to the values shown in Table 4, cluster 4 and cluster 1 could switch positions, while cluster 2 and cluster 3 are clearly separated.

In order to determine the differences between the identified clusters with respect to the other variables in the survey, we built three generalised linear models, one for comparing cluster 2 with cluster 3 (averse vs. proactive), a second one for comparing cluster 4 and cluster 1 (reluctant vs. open) and a third for comparing cluster 1 and cluster 3 (open vs. proactive). We have only included the control variables proven to be statistically significant in Table 4 ($p<0.01$) along their first-level interactions in order to see whether they are still relevant for separating elements of one cluster from the another, when all the relevant items in the survey are taken into account (Aibar et al., 2015).

Table 5 shows the reliability of the latent variables used to compute the generalised linear model and the explained variance by the (unique in all cases) factor computed using a non-rotated PCA. Notice that we are not trying to build a predictor for determining the most appropriate cluster for a given user, but analysing the contribution of each latent variable (and their interactions) to the fact of belonging to one given cluster or another, taking such contributions into account simultaneously. Thus, we have included all the latent variables in our analysis, even those with a low value for Cronbach's alpha ("Easiness" and "Visibility", $\alpha=0.58$ and $\alpha=0.56$, respectively), following the recommendations stated by Schmitt (1996). Furthermore, the percentage of explained variance is large enough to ensure that each latent variable captures most of the meaning of its original questionnaire items.

| Latent variable | Cronbach's α | Variance (%) | Mean (SD) |
| --- | --- | --- | --- |
| Profile 2.0 | 0.79 | 70.3 | 4.48 (1.89) |
| Experience | 0.75 | 62.2 | 6.00 (1.72) |
| Usefulness | 0.86 | 78.8 | 5.62 (1.56) |
| Easiness | 0.58 | 54.9 | 6.76 (1.08) |
| Quality | 0.86 | 70.0 | 6.31 (1.49) |
| Image | 0.71 | 63.5 | 5.00 (1.39) |
| Visibility | 0.56 | 56.5 | 4.14 (1.28) |
| Incentives | 0.84 | 67.4 | 7.04 (1.88) |

Table 5: Reliability and percentage of explained variance for the latent variables.

Table 6 shows the results for the first generalised model, comparing *averse* (represented by 0 in the generalised linear model as the outcome) and *proactive* users (represented by 1). For the sake of simplicity, we have omitted the non-statistically significant variables in Table 4. We have built two different versions: the model farthest to the left in Table 6 contains all significant variables, and a simplified version of this model without the control variables is also shown. We used an Akaike information criterion-based (AIC) stepwise algorithm in order to capture the true nature of faculty according to their attitudes and perceptions only, regardless of their gender or position.

Note that none of the control variables in Table 4 or any of their interactions appear in Table 6 (a), except "Gender" and its interaction with "Academia". All the items but "Gender" have a positive estimate, i.e. they show higher values for *proactive* users. "Visibility" is actually the most important variable, as its estimate has the largest impact (taking into account variable range), followed by "Experience". Note also the large value for the model intercept in both cases, which indicates that both clusters are very well separated. Notice also that the simplified model improves AIC while maintaining all latent variables, with proportional estimates.

|  | (a) Complete model with control variables | | | | (b) Simplified model without control variables | | | |
| --- | --- | --- | --- | --- | --- | --- | --- | --- |
|  | N=332, AIC=106.58, pseudo-$R^2$=0.87 | | | | N=332, AIC=88.59, pseudo-$R^2$=0.83 | | | |
| Variable | Estimate | Std. Error | z value | p | Estimate | Std. Error | z value | p |
| (Intercept) | -27.8661 | 6.3249 | -4.406 | < 0.001 | -23.5963 | 3.9750 | -5.936 | < 0.001 |
| Gender | -2.8145 | 1.2410 | -2.540 | 0.0233 | --- | --- | --- | --- |
| Academic | -1.3605 | 1.1317 | -1.202 | 0.2293 | --- | --- | --- | --- |
| Experience | 1.6003 | 0.4189 | 3.820 | < 0.001 | 1.2849 | 0.3018 | 4.257 | < 0.001 |
| Usefulness | 1.0861 | 0.4513 | 2.407 | 0.0161 | 0.5615 | 0.2749 | 2.043 | 0.0411 |
| Image | 1.1832 | 0.4200 | 2.817 | 0.0048 | 0.9262 | 0.3038 | 3.049 | 0.0023 |
| Visibility | 3.2217 | 0.7423 | 4.340 | < 0.001 | 2.6561 | 0.5039 | 5.271 | < 0.001 |
| Gender * Academic | 3.3380 | 1.8522 | 1.802 | 0.0715 | --- | --- | --- | --- |

**Table 6: Significant variables (p<0.1) for the generalised linear models (complete and simplified) separating clusters *averse* and *proactive*.**

The model described in Table 6 shows that faculty members would be classified as *averse* by default (due to this large negative intercept), that is, only those faculty members with larger values for the majority of the significant latent variables would be classified as *proactive*. In fact, these two clusters are very well separated.

A second generalised model for comparing *reluctant* (represented by 0) and *open* (represented by 1) clusters was also built. Our purpose here was to determine if there was any hidden trait or attitude that could be used to differentiate the users that were not *averse* or who were *proactive* users of Wikipedia. Once again, the model and its simplified version are shown. Table 7 only shows the variables that were found to be statistically significant for separating users from both clusters. Notice that *open* users displayed more positive attitudes and perceptions. In this case, all estimates are very similar, so no latent variable has a larger impact than the rest.

|  | (a) Complete model with control variables | | | | (b) Simplified model without control variables | | | |
| --- | --- | --- | --- | --- | --- | --- | --- | --- |
|  | N=314, AIC=364.27, pseudo-$R^2$=0.24 | | | | N=314, AIC=350.99, pseudo-$R^2$=0.20 | | | |
| Variable | Estimate | Std. Error | z value | p | Estimate | Std. Error | z value | p |
| (Intercept) | -8.1223 | 1.5751 | -5.157 | < 0.001 | -6.2998 | 1.1380 | -5.536 | < 0.001 |
| Usefulness | 0.4821 | 0.1367 | 3.527 | < 0.001 | 0.5309 | 0.1262 | 4.207 | < 0.001 |
| Quality | 0.3681 | 0.1381 | 2.665 | 0.0077 | 0.3709 | 0.1279 | 2.899 | 0.0038 |
| Image | 0.3036 | 0.1346 | 2.256 | 0.0241 | 0.2820 | 0.1233 | 2.287 | 0.0222 |
| Visibility | 0.3287 | 0.1702 | 1.931 | 0.0535 | 0.2540 | 0.1532 | 1.659 | 0.0972 |

**Table 7. Significant variables (p<0.1) for the generalised linear models (complete and simplified) separating clusters *reluctant* and *open*.**

Finally, a third model was built to discriminate between participants in the *open* and *proactive* categories, in order to see if there were some barriers that prevented someone from becoming a proactive Wikipedia user. Table 8 shows a more complex combination of factors than in the previous cases, as more variables appear in the model, and some of them with negative estimates. In this case, "Visibility" carries the most weight within this model, but the other latent variables also have comparable estimates.

|  | (a) Complete model with control variables | | | | (b) Simplified model without control variables | | | |
|  | N=367, AIC=419.92, pseudo-$R^2$=0.25 | | | | N=367, AIC=414.16, pseudo-$R^2$=0.21 | | | |
| Variable | Estimate | Std. Error | z value | p | Estimate | Std. error | z value | p |
| --- | --- | --- | --- | --- | --- | --- | --- | --- |
| (Intercept) | -4.8198 | 1.3486 | -3.574 | < 0.001 | -5.8880 | 1.0972 | -5.367 | < 0.001 |
| Gender | -0.8175 | 0.3977 | -2.056 | 0.0398 | --- | --- | --- | --- |
| Academic | -0.7173 | 0.4321 | -1.660 | 0.0969 | --- | --- | --- | --- |
| STEM | -0.2276 | 0.4592 | -0.495 | 0.6207 | --- | --- | --- | --- |
| Registered | 1.1041 | 0.6277 | 1.759 | 0.0786 | --- | --- | --- | --- |
| Experience | 0.3532 | 0.1163 | 3.037 | 0.0024 | 0.3774 | 0.1028 | 3.672 | < 0.001 |
| Usefulness | 0.2502 | 0.1227 | 2.039 | 0.0415 | 0.2993 | 0.1098 | 2.726 | 0.0064 |
| Easiness | -0.3433 | 0.1364 | -2.517 | 0.0118 | -0.2908 | 0.1271 | -2.289 | 0.0221 |
| Visibility | 0.6779 | 0.1388 | 4.884 | < 0.001 | 0.7193 | 0.1219 | 5.901 | < 0.001 |
| Gender * STEM | 1.5932 | 0.7972 | 1.998 | 0.0457 | --- | --- | --- | --- |

Table 8. Significant variables (p<0.1) for the generalised linear models (complete and simplified) separating "proactive" and "open" clusters.

**Discussion**

In the light of the results obtained in the previous section, we can proceed with the characterisation of the typical profile represented by each cluster. First, we establish differences between the two opposite clusters 2 and 3, i.e. between *averse* and *proactive* users according to their use of Wikipedia as a tool for teaching. We then characterise users in the other two clusters, ending with a discussion about the relevance of the rest of variables described in Table 4.

In summary, faculty classified as *proactive* (i.e. falling into cluster 3) are mostly men, part-time faculty, not necessarily with a PhD (which, in general, is highly correlated with the previous fact, r=0.77, N=905), who create and share open resources. Most of them are teaching in STEM-related subjects, especially engineering. They use Wikipedia for academic and personal purposes and an important percentage are also registered Wikipedia users who contribute as well. They also cite Wikipedia articles in their texts, as they do not consider the quality of Wikipedia articles to be lower than classical information sources in their field of expertise.

Therefore, this is a significant group of faculty members that use Wikipedia as more than a mere information source (Knight and Pryke, 2012). This group is consistent with the results obtained by H. Chen (2010), showing a high correlation between using Wikipedia as an information source and using it for teaching or research purposes. We can also find some evidence of these *proactive* users in the interviews from our exploratory qualitative study, for example:

- [Interview 1]: '*[I edit because of] the quirk of seeing a mistake and want to correct it.*'
- [Interview 1]: '*I agree that students use it if they know how to use Wikipedia, that is to say, if they have some skills in critical reading and if they can, for example, check the sources, make sure that references are being cited, if they are able to check the*

*history page, if there have been controversies, etc.; so, yes, I totally agree that students use it.*

On the contrary, faculty classified as *averse* (cluster 2) are mostly women (by a small although significant proportion), part-time or full-time academic faculty in non-STEM subjects, half of them with a PhD. They do not create or share open resources and they are not registered Wikipedia users or contributors, showing a medium use of Wikipedia for personal purposes. They never or rarely cite Wikipedia articles in their texts, as they strongly agree with the idea that the quality of Wikipedia articles is lower than classical information sources, especially with respect to completeness.

The behaviour and attitudes found in this cluster are consistent with the negative perception of Wikipedia described in (Francke and Sundin, 2010; Kubiszewski et al., 2011), which could be partially supported as Wikipedia has recently received negative media publicity (Cohen, 2007; Waters, 2007; Maehre, 2009). The influence of gender described in (Lim and Kwon, 2010) is also consistent with the differences found between proactive and averse users, as males seem to enjoy more from Wikipedia than females, regardless the perceived credibility. Some examples of an *averse* user from our interview set would be:

- [Interview 10]: *'I never cite Wikipedia in my scholarship because the use I make of it [Wikipedia] never ends in a reference category.'*
- [Interview 10]: *'It bothers me [that students cite Wikipedia in their assignments]. It bothers me a lot. In the same way that it would bother me if they do their assignments according to the Catalan Encyclopaedia. Academic assignments are not meant to be done on the basis of encyclopaedic articles.'*

As mentioned before, these two clusters are very well separated. Among the items in the survey found to be significant in Table 6, citing Wikipedia in their academic works (i.e. "Visibility") is one of the most important, showing that *proactive* faculty not only embrace Wikipedia but are also not afraid to acknowledge that they do.

A high use of Wikipedia for both personal and professional purposes (i.e. "Experience") is also a good indicator of its use as a teaching tool. However, using other Web 2.0 tools such as blogs or social networking tools is not significant, which was found to positively correlate with using Web 2.0 tools in online courses (Ulrich and Karvonen, 2011). Hence, it should not be very difficult for institutions to identify an incipient set of Wikipedia champions that could help other faculty members to use Wikipedia, leaving behind any prejudices and misconceptions they may have about it and adopting and spreading the ideas described in Konieczny (2012). As stated by Smith (2012), innovations championed by practitioners within a discipline are more likely to be successful within that discipline than if the champions are centrally based. Therefore, receiving training from a colleague can be a driver for disseminating the use of Wikipedia among faculty from the same department.

The other two clusters (*reluctant* and *open*) show no particular traits when only control variables are considered. Nevertheless, there are several differences according to Table 7 that partially explain why we found two different clusters instead of just one. A majority of faculty do not use Wikipedia as a teaching tool, but they are not against its use, mostly falling into the *open* cluster:

- [Interview 2]: *'I do consult specialized books, etc., for technical things. But in order to have the first idea of a topic, I do not consult anything but Wikipedia.'*
- [Interview 7]: *'Yes, I agree that students use it (...) within a context (...). I think it [has become] a part of our common informational environment, so, we have to accept it as such.'*

As expected, *open* users have a better opinion about the quality of Wikipedia articles than *reluctant* users. Furthermore, *open* users also think that Wikipedia is useful and that it is used and viewed well by their colleagues. Thus, there must be some barriers that prevent these users from proactively using Wikipedia as a teaching tool. Although "Incentives" is not a significant variable (thus not shown in Table 7) by a small amount (p=0.1121), it deserves special attention, as it might be useful in identifying these possible barriers. In fact, both INC2 (having a colleague explaining their experience) and INC3 (receiving specific training) show higher values for *open* users (p<0.001 and p=0.0159, respectively). This is consistent with the results described by Wannemacher (2011), where support is identified as a need to facilitate course work with Wikipedia.

This could be understood as a real need of *reluctant* faculty, who might think they need to know more about Wikipedia and would feel much better if such knowledge came from a colleague, thus reinforcing the idea that using Wikipedia is supported by academia. Accordingly, *proactive* users within the institution could act as Wikipedia ambassadors, advocating its adoption as a teaching tool, but from a more reputable position. Xiao and Askin (2014) show that researchers' academic environment seems to have some impact on their opinions about Wikipedia publishing, and the more experiences with Wikipedia publishing they have, the more positive the academic researcher's attitudes towards it, suggesting that promoting their participation in Wikipedia can improve their perspectives on the idea.

Finally, when comparing *open* and *proactive* users, there are several differences according to the models described in Table 8. Our goal is to identify the ultimate barriers that an *open* faculty member must overcome in order to adopt Wikipedia as a teaching tool. Besides the smaller values of "Experience", "Usefulness" and "Visibility" for *open* faculty, the most interesting fact revealed by Table 8 is that *open* faculty members have a more negative perception of Wikipedia's ease of use than *proactive* users.

This fact could be one of the barriers to faculty being willing to use Wikipedia as a teaching tool – they think they lack the specific skills to do so. Furthermore, as also shown in Table 8, quality is not an issue for *open* faculty, as they do not tend to think of Wikipedia as a poorer information source. Therefore, *open* faculty members might just need some training (once again, maybe from institutional Wikipedia ambassadors) in order to become part of the *proactive* cluster. As abovementioned, these ambassadors should be from the same academic discipline (Smith, 2012).

Regarding the control variables, in general, our analysis also shows that age is not an important variable in determining the attitude towards using Wikipedia as a tool in the classroom. As shown in Table 4, average age in each cluster is approximately the same. It would seem that the usual association between younger people and use of Web 2.0 tools (including social media) does not work in this case. Teaching expertise measured using Huberman's scale (1993) is not relevant either, which is not surprising, since teaching expertise is heavily associated with age. As shown in Table 4, being an academic faculty member or an adjunct professor is significant (p=0.0069), although this difference is only important between *proactive* faculty members and the rest, as *proactive* faculty are mostly part-time. At the same time, holding a PhD is only significant at a 0.1 confidence level (p=0.0752), as most adjunct faculty working part-time do not satisfy such a requirement, while most full-time faculty do.

On the other hand, regarding academic disciplines, our results show that there are significant differences among clusters, especially in Engineering, as described by H. Chen (2010). As shown in Table 4, STEM faculty are underrepresented in cluster 2 (*averse*) and overrepresented in cluster 3 (*proactive*). Table 6 also shows that non-STEM faculty have a more negative perception of the quality of Wikipedia's articles, which is consistent with the overall quality differences across scientific disciplines, as described by Ehmann et al. (2008). If faculty from non-STEM fields never engage in using (and improving) Wikipedia, they will continue to perceive it as a lower-quality resource, and the situation will be perpetuated. As stated by Smith et al. (2008), STEM faculty is usually research-oriented, and more amenable to collaborative teaching, while non-STEM faculty usually less competitive (with respect to research) and less likely to share collaborative teaching practices (as scholarly knowledge in this area has a more individualistic nature). Furthermore, applied disciplines (such as Engineering) are moving in the direction of diversification and community practice, directing students to new sources of information, being Wikipedia the one with highest online visibility. Teachers in hard disciplines (i.e. STEM) are more likely to focus their teaching in transmitting information (Stes and Van Petegem, 2014).

**Conclusions**

Wikipedia is one of the most successful websites, accessed by millions of users every day. It has become a reliable source of information, but higher education faculty are still averse to incorporating the use of Wikipedia as another teaching tool in the classroom. This could be due to a combination of prejudices and lack of knowledge about how Wikipedia works and what the real possibilities for using it are. In this paper we have described a survey launched to determine such causes, which was followed by a clustering analysis, the aim of which was to determine whether there are different usage profiles among faculty or not and identify the variables that best describe such differences among clusters. We concluded by discussing the generalised linear logit models built in order to quantify the importance of each variable with respect to belonging to one cluster or another.

Our analysis has produced four different clusters, named *averse*, *reluctant*, *open* and *proactive*, which were sorted according to Wikipedia usage for teaching purposes. Our first finding is that, contrary to the standard view, a majority of faculty does not share the most negative stand (represented by our *averse* cluster). In fact, the more positive positions (*proactive* and *open*) gather more than the half of the respondents (56.7 %).

We have also shown that there are very significant differences between the two most separated clusters. On the one hand, the first cluster, named *proactive*, comprises mostly part-time male faculty in STEM fields. These faculty members are fully engaged in using Wikipedia for both personal and academic purposes and they are not afraid of sharing this fact with their colleagues. On the other hand, an opposite cluster, named *averse*, comprises mostly full-time female faculty in non-STEM fields. They do not use Wikipedia and they think their colleagues do not use it either.

Between these two clusters we found two other clusters composed of people that do not use Wikipedia as a teaching tool but take a different stance in regard to their opinion about using it, as their *reluctant* and *open* colleagues do. The latter would be more predisposed to using Wikipedia if they had some positive examples and possibly training from colleagues who already use it, as this would help them to overcome certain prejudices related to its use in an academic context.

It is remarkable that control variables such as age, having a PhD or level of teaching experience do not appear to be important when deciding to use Wikipedia as a teaching tool. Therefore, perceptions and attitudes caused by environmental factors (i.e. the academic culture and the scientific discipline) or their lack of knowledge, misconceptions and prejudices (Cohen, 2007; Eijkman, 2010; Knight and Pryke, 2012; Ricaurte-Quijano and Alvarez, 2016) are the underlying reasons for not using it.

Higher education institutions can take advantage of Wikipedia as a huge source of quality information and, at the same time, improve it by helping their faculty become aware of the possibilities of using Wikipedia as a teaching tool. They can detect Wikipedia champions within each institution and help them to become Wikipedia ambassadors who share their knowledge with their colleagues, thus creating a spillover effect (Ricaurte-Quijano and Alvarez, 2016). This would promote a win-win situation for both Wikipedia and higher education faculty, transferring knowledge from academia to the general public that uses Wikipedia as their primary source of scientific information while increasing faculty visibility and opportunities for improving their teaching practices.

Further research into this topic should handle both the limitations and main findings of this study. Repeating the survey with more faculty from traditional brick-and-mortar universities will help to better measure the impact that academic discipline has on using Wikipedia or not, as there are probably significant differences between STEM and non-STEM fields. The importance of the gender variable also deserves further investigation to determine the existence of hidden attitudes towards Wikipedia's collaborative editing process, especially for non-STEM fields where women are in the majority among faculty. Furthermore, the design, implementation and evaluation of policies and projects promoting the use of Wikipedia among faculty will also be useful in determining present barriers for each of the four profiles found in this study.

## References

Aibar, E., Lladós, J., Meseguer-Artola, A., Minguillón, J., & Lerga, M. (2015). Wikipedia at university: What faculty think and do about it. *The Electronic Library, 33*(4), 668-683.

An, Y.-J. and Williams, K. (2010). Teaching with Web 2.0 Technologies: Benefits, Barriers and Lessons Learned. *International Journal of Instructional Technology & Distance Learning, 7*(3), article 4.

Bayliss, G. (2013). Exploring the cautionary attitude toward Wikipedia in higher education: implications for higher education institutions. *New Review of Academic Librarianship, 19*(1), 36-57.

Benkler, Y. (2006). The Wealth of Networks: How Social Production Transforms Markets and Freedom. Yale University Press, New Haven, CT, USA.

Black, E.W. (2008). Wikipedia and academic peer-review – Wikipedia as a recognized medium for scholarly publication?. *Online Information Review, 32*(1), 73-88.

Borg, I., Groenen, P. (2005). Modern Multidimensional Scaling: theory and applications. (2nd ed.) (pp. 207–212). New York: Springer-Verlag.

Brossard, D., & Scheufele, D.A. (2013). Science, New Media, and the Public. *Science, 339*(6115), 40–41.

Brox, H. (2012). The Elephant in the Room: a Place for Wikipedia in Higher Education?, *Nordlit, 30*, 143−155.

Caliński, T. and Harabasz, J. (1974). A dendrite method for cluster analysis. *Communications in Statistics-Simulation and Computation, 3*, 1-27.

Chandler, C. J., & Gregory, A. S. (2010). Sleeping with the Enemy: Wikipedia in the College Classroom. *History Teacher, 43*(2), 247-257.

Chen, H. (2010). The perspectives of higher education faculty on Wikipedia. *The Electronic Library, 28*(3), 361-373.

Chen, S. (2010). Wikipedia: a Republic of Science Democratized. *Albany Law Journal of Science and Technology, 20*(2), 249-326.

Christensen, T.B. (2015). Wikipedia as a Tool for 21st Century Teaching and Learning. *International Journal for Digital Society, 6*(2), 1042-1047.

Cohen, N. (2007). A history department bans citing Wikipedia as a research source. The New York Times. New York, The New York Times Company. [Available at http://www.nytimes.com/2007/02/21/education/21wikipedia.html]

Dooley, P. (2010). "Wikipedia and the two-faced professoriate", in Wikisym'10 Proceedings of the 16th International Symposium on Wikis and Open Collaboration in Gdansk, Poland 2007, Article No. 24, ACM, New York, NY.

Ehmann, K., Large, A., Behesti, J. (2008). Collaboration in context: comparing article evolution among subject disciplines in Wikipedia. *First Monday, 13*(10). [Available at http://firstmonday.org/ojs/index.php/fm/article/view/2217/2034]

Eijkman, H. (2010). Academics and Wikipedia: Reframing Web 2.0+ as a disruptor of traditional academic power-knowledge arrangements. *Campus-Wide Information Systems, 27*(3), 173-185.

Forte, A., Bruckman, A. (2006). From Wikipedia to the classroom: Exploring online publication and learning. In Proceedings of the 7th international conference on Learning sciences (pp. 182-188). International Society of the Learning Sciences.

Francke, H. & Sundin, O. (2010). An inside view: Credibility in Wikipedia from the perspective of editors. *Information Research, 15*(3), paper colis702. [Available at http://InformationR.net/ir/15-3/colis7/colis702.html]

Gower, J. C. (1971). A general coefficient of similarity and some of its properties. *Biometrics, 27*, 857–871.

Gray, K., Thompson, C., Cerehan, R., Sheard, J., Hamilton, M. (2008). Web 2.0 authorship: Issues of referencing and citation for academic integrity. *The Internet and Higher Education, 11*(2), 112-118.

Hair, J.F., Black, W.C., Babin, B.J., Anderson, R.E. (2010). Multivariate data analysis, 7th ed., Prentice Hall, Upper Saddle River, NJ.

Halfaker, A. and Taraborelli, D. 2015. Scholarly article citations in Wikipedia. Figshare. [Available at http://dx.doi.org/10.6084/m9.figshare.1299540]

Huberman, M. (1993). Steps toward a developmental model of the teaching career. In L. Kremer-Hayon, H. C. Vonk, & R. Fessler (Eds.), Teacher professional development: A multiple perspective approach (pp. 93–118). Amsterdam: Swets & Zeitlinger.

Jashchick, S. (2007). A stand against Wikipedia. Inside Higher Education, 26. [Available at https://www.insidehighered.com/news/2007/01/26/wiki]

Kaufman, L., Rousseeuw, P.J. (1987). Clustering by means of Medoids, in Statistical Data Analysis Based on the L1–Norm and Related Methods, edited by Y. Dodge (pp. 405-416). North-Holland.

Kemp, B., Jones, C. (2007). Academic use of digital resources: Disciplinary differences and the issue of progression revisited. *Educational Technology and Society, 10*(1), 52–60.

Kennedy, R., Forbush, E., Keegan, B., & Lazer, D. (2015). Turning introductory comparative politics and elections courses into social science research communities using wikipedia: Improving both teaching and research. *PS, Political Science & Politics, 48*(2), 378-384.

Knight, C., Pryke, S. (2012). Wikipedia and the University, a case study. *Teaching in Higher Education, 17*(6), 649-659.

Konieczny, P. (2007). Wikis and Wikipedia as a teaching tool. *International Journal of Instructional Technology & Distance Learning, 4*(1).

Konieczny, P. (2012). Wikis and Wikipedia as a teaching tool: Five years later. *First Monday, 17*(9). [Available at http://firstmonday.org/article/view/3583/3313]

Konieczny, P. (2014). Rethinking Wikipedia for the classroom. *Contexts, 13*(1), 80-83.

Konieczny, P. (2016). Teaching with Wikipedia in a 21st-Century Classroom: Perceptions of Wikipedia and Its Educational Benefits. *Journal of the Association for Information Science and Technology, 67*(7), 1523-1534.

Kubiszewski, I., Noordewier, T., Constanza, R. (2011). Perceived credibility of Internet encyclopedias. *Computers & Education, 56*(3), 659-667.

Lim, S. (2009). How and why do college students use Wikipedia?, *Journal of the American Society for the Information Science and Technology, 60*(11), 2189-2222.

Lim, S., Kwon, N. (2010). Gender differences in information behavior concerning Wikipedia, an unorthodox information source?. *Library & Information Science Research, 32*(3), 212-220.

Maehre, J. (2009). What it means to ban Wikipedia: an exploration of the pedagogical principles at stake. *College Teaching, 57*(4), 229-236.

Meseguer-Artola, A. (2014). Learning by comparing with Wikipedia: the value to students' learning. *RUSC. Universities and Knowledge Society Journal, 11*(2), 57-69.

Meseguer-Artola, A., Aibar, E., Lladós, J., Minguillón, J., & Lerga, M. (2016). Factors that influence the teaching use of Wikipedia in higher education. *Journal of the Association for Information Science and Technology, 67*(5), 1224-1232.

Nielsen, F. A. (2007). Scientific Citations in Wikipedia. *First Monday 12*(8). [Available at http://firstmonday.org/ojs/index.php/fm/article/view/1997]

Nix, E. M. (2010), Wikipedia: How It Works and How It Can Work for You. *History Teacher, 43*(2), 259-264.

Okoli, C., Mesgari, M., Mehdi, M., Nielsen, F.A., Lanamaki, A. (2014). Wikipedia in the Eyes of Its Beholders: A Systematic Review of Scholarly Research on Wikipedia Readers and Readership. *Journal of the Association for Information Science and Technology, 65*(12), 2381-2403.

Ricaurte-Quijano, P., Alvarez, A.C. (2016). The Wiki Learning Project: Wikipedia as an Open Learning Environment. *Comunicar, 49*, 61-69.

Samoilenko, A., Yasseri, T. (2014). The distorted mirror of Wikipedia: a quantitative analysis of Wikipedia coverage of academics. *EPJ Data Science, 3*(1), 1-11. [Available at http://epjdatascience.springeropen.com/articles/10.1140/epjds20]

Santana, A., Wood, D. J. (2009). Transparency and social responsibility issues for wikipedia. *Ethics and Information Technology, 11*(2), 133-144.

Schmitt, N. (1996). Uses and abuses of coefficient alpha. *Psychological Assessment, 8*(4), 350-353.

Smith, G.G., Heindel, A.J., Torres-Ayala, A.T. (2008). E-learning commodity or community: Disciplinary differences between online courses. *Internet and Higher Education, 11*(3-4), 152-159.

Smith, K. (2012). Lessons learnt from literature on the diffusion of innovative learning and teaching practices in higher education. *Innovations in Education and Teaching International, 49*(2), 173-182.

Snyder, J. (2010). Wikipedia as an Academic Reference: Faculty and Student Viewpoints. In Proceedings of the 16th Americas Conference on Information Systems, Paper 17.

Stes, A., Van Petegem, P. (2014). Profiling approaches to teaching in higher education: a cluster-analytic study. *Studies in Higher Education, 39*(4), 644-658.

Soules, A. (2015). Faculty Perception of Wikipedia in the California State University System. *New Library World, 116*(3/4), 213-226.

Traphagan, T., Traphagan, J., Neavel Dickens, L., & Resta, P. (2014). Changes in college students' perceptions of use of web-based resources for academic tasks with Wikipedia projects: a preliminary exploration. *Interactive Learning Environments, 22*(3), 253-270.

Ulrich, J., Karvonen, M. (2011). Faculty instructional attitudes, interest and intention: Predictors of Web 2.0 use in online courses. *The Internet and Higher Education, 14*(4), 207-216.

Wannemacher, K. (2009). Articles as Assignments - Modalities and Experiences of Wikipedia Use in University Courses. In Advances in Web Based Learning–ICWL 2009 (pp. 434-443). Springer Berlin Heidelberg.

Wannemacher, K. (2011). Experiences and perspectives of Wikipedia use in Higher Education. *International Journal of Management in Education, 5*(1), 79-92.

Waters, N.L. (2007). Why you can't cite Wikipedia in my class. *Communications of the ACM, 50*(9), 15–17.

Weber, S. (2004). The Success of Open Source. Harvard University Press: Cambridge, MA.

Xiao, L., Askin, N. (2012). Wikipedia for academic publishing: advantages and challenges. *Online Information Review, 36*(3), 359-373.

Xiao, L., Askin, N. (2014). Academic opinions of Wikipedia and Open Access publishing. *Online Information Review, 38*(3), 332-347.